\documentclass[useAMS]{mn2e}
\usepackage{times}
\usepackage{amssymb}
\usepackage{rotating}
\input{epsf}

\title[Physical PSD model]
{Modelling variability in black hole binaries:
linking simulations to observations}
\author[A. Ingram \& Chris Done]
{Adam
Ingram$^{1}\thanks{E-mail:a.r.ingram@durham.ac.uk}$ \&
Chris Done$^{1}$\\
$^1$Department of Physics, University of Durham, South Road,
Durham DH1 3LE, UK\\
}

\date{Submitted to MNRAS}
\pagerange{\pageref{firstpage}--\pageref{lastpage}} \pubyear{2010}

\begin{document}

\topmargin = -0.5cm

\maketitle

\label{firstpage}

\begin{abstract}

Black hole accretion flows show rapid X-ray variability. The Power
Spectral Density (PSD) of this is typically fit by a phenomenological
model of multiple Lorentzians for both the broad band noise and
Quasi-Periodic Oscillations (QPOs). Our previous paper (Ingram \& Done
2011) developed the first physical model for the PSD and fit this to
observational data. This was based on the same truncated disc/hot
inner flow geometry which can explain the correlated properties of the
energy spectra. This assumes that the broad band noise is from
propagating fluctuations in mass accretion rate within the hot flow,
while the QPO is produced by global Lense-Thirring precession of the
same hot flow.

Here we develop this model, making some significant improvements.
Firstly we specify that the viscous frequency (equivalently, surface
density) in the hot flow has the same form as that measured from
numerical simulations of precessing, tilted accretion flows.
Secondly, we refine the statistical techniques which we use to fit the
model to the data.  We re-analyse the PSD from the 1998 rise to
outburst of XTE J1550-564 with our new model in order to assess the
impact of these changes. We find that the derived outer radii of the
hot flow (set by the inner radius of the truncated disc) are rather
similar, changing from $\sim 68-13 R_g$ throughout the outburst
rise. However, the more physical assumptions of our new model also
allow us to constrain the scale height of the flow. This decreases as
the outer radius of the flow decreases, as expected from the spectral
evolution. The spectrum steepens in response to the increased cooling
as the as the truncation radius sweeps in, so gas pressure support for
the flow decreases.

The new model, \textsc{propfluc}, is publically available within the
{\sc xspec} spectral fitting package.  

\end{abstract}


\section{Introduction} \label{sec:introduction}

Black Hole Binaries (BHBs) have X-ray emission which is variable on a
broad range of timescales. On the longest timescales ($\sim$ weeks),
these sources are seen to transition between quiescence, when they are
hardly visible above the X-ray background; and outburst, when they are
amongst the brightest X-ray objects in the sky. During the rise from
quiescence to outburst, the source displays an evolution in spectral
state. At the lowest luminosities, the source is in the low/hard
state, with a Spectral Energy Distribution (SED) dominated by a hard
(photon index $\Gamma < 2$) power law tail but also including a weak
disc component and reflection features.  As the average mass accretion
rate, $\dot{M}_0$, increases, the disc increases in luminosity, the
power law softens ($\Gamma \sim 2$) and the reflection fraction
increases (intermediate state). As the luminosity increases further,
the SED either becomes almost completely disc dominated (high/soft or
ultrasoft state) or it displays both a strong disc component with a
soft ($\Gamma>2$) power law tail (very high state; see e.g Done,
Gierlinski \& Kubota 2007, hereafter DGK07).

The SED of the low/hard state can be explained by a two component
model whereby the standard cool, optically thick, geometrically thin
accretion disc (Shakura \& Sunyaev 1973) is truncated at some radius
$r_o$ which is greater than the last stable orbit, $r_{lso}$. Interior
to this is a hot, optically thin, geometrically thick accretion flow,
perhaps similar to an Advection Dominated Accretion Flow (ADAF;
Narayan \& Yi 1995). We observe a fraction of the disc emission
directly but some of the photons emitted by the disc are incident upon
the flow where they are Compton up scattered by the hot electrons thus
creating the power law tail. As the source evolves, the truncation
radius decreases which increases both the amount of direct disc
emission and the number of seed photons incident on the flow thus
cooling the Comptonizing electrons and softening the power law index
of the flow emission. The spectral evolution throughout the rise to
outburst can be described by assuming that $r_o$ decreases as the
average mass accretion rate increases until $r_o \approx r_{lso}$ in
the high/soft state (Esin, McClintock \& Narayan 1997, DGK07). 
While this truncated disc model has been challenged by direct observation of
the disc inner edge at the last stable orbit in the low/hard state
(Miller, Homan \& Miniutti 2006; Rykoff et al 2007), this interpretation of the
data has itself been challenged (Done \& Diaz Trigo 2010; Gierlinski,
Done \& Page 2009). Thus there is no unambiguous evidence as yet which
rules out these models, and all alternative geometries (e.g. Markoff,
Nowak \& Wilms 2005; Miller, Homan \& Miniutti 2006) run into other
difficulties (DGK07). Hence we use the truncated disc model as our
overall framework, and explore how this can also be used to 
interpret the rapid variability seen within a single observation
of $\sim 2000s$. 

A power spectral analysis of this rapid variability reveals
Quasi-Periodic Oscillations (QPOs) superimposed on a broad band noise
of variability. The broad band noise can be described using broad
Lorentzians centred at characteristic frequencies $f_b$ and $f_h$,
often referred to as the low and high frequency breaks. The QPO
(fundamental and higher harmonics) can be described by narrow Lorentzians
centred at $f_{QPO}$, $2f_{QPO}$ etc (e.g. Belloni, Psaltis \& van der
Klis 2002). As the SED evolves on long timescales, so does the
corresponding power spectral density (PSD).  All the PSD frequencies,
$f_b$, $f_{QPO}$ and (to a lesser extent) $f_h$ increase with
luminosity and are correlated (Wijnands \& van der Klis 1999;
Klein-Wolt \& van der Klis 2008; van der Klis 2004; Psaltis, Belloni
\& van der Klis 1999; Belloni 2010). However, the amount of power at high
frequencies ($>10~Hz$), remains constant despite the increase in $f_h$
(Gierli{\'n}ski, Niko{\l}ajuk, \& Czerny 2008). Although the
variability properties have been studied for over 20 years, the
underlying physical processes are still poorly understood. Most
previous attempts to physically model the variability properties focus on
explaining either the QPO or the broadband noise. We review these
briefly below. 

\subsection{QPO models}

There are many proposed QPO mechanisms in the literature, some of
which are based on a misalignment between the angular momentum of the
black hole and that of the binary system (e.g. Stella \& Vietri 1998;
Fragile, Mathews \& Wilson 2001; Schnittman 2005; Schnittman et al
2006; Ingram, Done \& Fragile 2009, hereafter IDF09) and others are
associated with wave modes in the accretion flow (Wagoner et al 2001;
Titarchuk \& Oscherovich 1999; Cabanac et al 2010). In IDF09, we
outlined a QPO model based on the original relativistic precession
model of Stella \& Vietri (1998). Here, the QPO frequency is given by
the Lense-Thirring precession frequency of a test mass (say, a hot
spot; e.g. Schnittman 2005) at the truncation radius.  Lense-Thirring
precession is a relativistic effect which occurs because of the
asymmetric gravitational potential present around a spinning black
hole. A test mass orbit will precess around the black hole if it is
misaligned with the black hole spin plane as spacetime is being
dragged around the black hole. As the particle reaches the starting
point of its orbit (i.e. $\phi =0, 2\pi, etc$), that point in
spacetime has rotated some way around the black hole.

Our extension of this model was to replace the test mass with the
entire hot accretion flow interior to the disc truncation radius.
There is a differential warp across the flow due to the radial
dependence of Lense-Thirring precession ($f_{LT}\propto \sim r^{-3}$).
Warps are communicated by bending waves which travel at (close to) the
sound speed. For a hot flow the sound speed is fast, so the warp can
be communicated across the whole hot inner flow on a timescale which
is \textit{shorter} than the precession period on the outer edge.  The
entire hot flow can then precess as a solid body (though it is still
differentially rotating), with a precession frequency given by the
surface density weighted average of $f_{LT}(r)$. This global
precession is confirmed by General Relativistic Magnetohydrodynamic
(GRMHD) simulations of a tilted flow (Fragile et al 2007, 2009). A
cool, thin disc responds very differently as here the sound speed
across the disc is much longer than the precession period. Viscous
diffusion then results in a steady state warp of the disc into the
plane of the black hole spin at small radii (Bardeen \& Petterson
1975; Kumar \& Pringle 1985, Fragile, Mathews \& Wilson 2001), rather
than global precession about the black hole spin axis.

In IDF09, we showed how global precession of the entire hot inner flow
interior to a stationary truncated disc can give the QPO. The
increasing frequency can be produced by the outer radius of the hot
flow (set by the inner radius of the truncated disc) moving inwards
from $50 > r_o >10$, as also implied by the SED evolution. This model
is also attractive as it ties the QPO to the hot flow, so trivially
modulates the Comptonised spectrum rather than the disc component, as
required by the data (Rodriguez et al 2004; Sobolewska \& Zycki 2005).

\subsection{Broad band noise models}

The underlying viscosity mechanism in the flow is most likely the
Magneto Rotational Instability (MRI; Hawley \& Balbus 1991). It is
looking increasingly likely that this is also the underlying source of
broad band variability in the flow as it generates large fluctuations
in all quantities (e.g. Krolik \& Hawley 2002; Dexter \& Fragile 2011;
Beckwith, Hawley \& Krolik 2008).
The temporal variability generated by the MRI extends to very high
frequencies but the emission is inherently linked to the reservoir of available
gravitational energy and therefore should depend on the mass accretion
rate, $\dot{M}$. It is commonly assumed that the variability in
$\dot{M}$ at a given radius is characterised by the local viscous
frequency, $f_{visc}(r)$, where $f_{visc}(r)\propto r^{-3/2}$ in the
simplest case. This interpretation implies that $f_b\approx
f_{visc}(r_o)$ and $f_h\approx f_{visc}(r_i)$ where $r_i$ is the inner
radius of the flow (e.g. Ingram \& Done 2010). Although this is a
rather crude approximation, it has the very attractive property that
the low frequency noise is tied to the truncation radius but the high
frequency noise is not. This can explain the observed correlation of
the low frequency break in the PSD with the low frequency QPO, as both
are set by $r_o$, while the high frequency break is more or less
constant as $r_i$ does not change(Gierli{\'n}ski, Niko{\l}ajuk, \&
Czerny 2008).

Another fundamental property of the data which must be reproduced by
any variability model is the sigma-flux relation (Uttley \& McHardy
2001). This can be measured by splitting the light curve into multiple
short segments and finding the average and standard deviation of each
segment. After binning, the standard deviation is always seen to be
linearly related to the average flux. This shows that the variability
is correlated across all timescales and therefore, in a picture where
different temporal frequencies come predominantly from different
spatial regions, there must be a causal connection between those
regions. This rules out simple shot noise models, where the
variability is independent. Instead, the sigma-flux relation
\textit{can} be reproduced if the fluctuations in $\dot{M}$ generated
at a given radius \textit{propagate} inwards towards the black hole as
might be intuitively expected in an accretion flow (Lyubarskii 1997;
Kotov et al 2001; Arevalo \& Uttley 2006 hereafter AU06)

\subsection{Combining the QPO and broad band noise}

In Ingram \& Done 2011 (hereafter ID11), we explored combining these
ideas of both QPO and noise by developing a model with propagating
mass accretion rate fluctuations in a precessing flow. The mass
accretion rate fluctuations produce the observed band limited noise
and the precession frequency modulates the fluctuating light curve to
create the QPO. Crucially, these two processes are linked. We use the
same geometry (inner and outer radius of the hot flow, surface density
of the hot flow) to make both the QPO {\sl and} broadband noise
properties. Precession of the fluctuating flow modulates its observed
emission, imprinting the QPO on the broadband noise, while
fluctuations in the flow cause fluctuations in the precession
frequency, making a quasi-periodic rather than periodic oscillation.

Here we develop a more advanced version of the model which is in
better agreement with the results of GRMHD simulations, together with
better statistical techniques to fit these models to the PSD data. We
have made this package publically available within {\sc xspec} (Arnaud
et al 1996) as a local model, {\sc propfluc}.

\section{The model}
\label{sec:mod}

As in ID11, the model consists of fluctuations in mass accretion rate
which propagate towards the black hole (following Lyubarskii 1997,
Kotov et al 2001 and AU06) within a flow that is precessing.  Here we
develop the model to include a number of improvements which allow us
to gain more physical insight from the best fit parameters. Most
significantly, we change our underlying assumption about the viscous
frequency $f_{visc}(r)$. In ID11 we assumed that this was a power law
between $r_i$ and $r_o$, the inner and outer radius of the precessing
hot flow. Here we have it be a smoothly broken power law, with the
radius of the break being the bending wave radius, $r_{bw}$, expected
from a misaligned flow. The viscous frequency is related to the
surface density profile, $\Sigma$ via the radial infall velocity
$v_r(r)$ as $f_{visc}(r)=-v_r(r)/R$ and mass conservation sets
$\dot{M}\propto \Sigma 2\pi r v_r$. Hence we can use the surface
density profiles from the GRMHD simulations to derive
$f_{visc}(r)$, which is especially important as the QPO frequency is
dependent on $\Sigma(r)$. 

We also change the assumed emissivity from ID11, where
$\epsilon \propto r^{-\gamma}b(r)$ (where $b(r)$ was an unknown
boundary condition) to $\epsilon \propto r^{-\gamma}\Sigma(r)$ i.e. we 
tie the emission to where the mass is in the flow.
We describe the details of the model below, mainly focusing
on these improvements made since ID11. Note that, throughout
the paper, we use the convention $R=rR_g$ where $R_g=GM/c^2$
is a gravitational radius and we always assume a 10 solar mass
black hole.

\subsection{Steady state properties}
\label{sec:sigma}

The surface density of the flow sets the QPO frequency by global
precession as
\begin{equation}
f_{prec} = \frac{\int_{r_{i}}^{r_{o}} f_{LT}f_{k}\Sigma r^3 dr}
{\int_{r_{i}}^{r_{o}} f_{k}\Sigma r^3 dr}
\label{eqn:fprec}
\end{equation}
where $f_k$ is the Keplerian orbital frequency and $f_{LT}$ is the point
particle Lense-Thirring precession frequency (given by equation 2 in ID11).

We use the GRMHD simulations of tilted flows to guide our description
of $\Sigma(r)$ (Fragile et al 2007; 2009; Fragile 2009).  These can be well 
fit by a smoothly broken power law function
\begin{equation}
\overline{\Sigma(r_n,t)}= \frac{\Sigma_0 \dot{M}_0}{cR_g}
\frac{x^\lambda}{(1+x^\kappa)^{(\zeta+\lambda)/\kappa}}.
\label{eqn:sigmabar}
\end{equation}
(IDF09), where $x=r/r_{bw}$ is radius normalised to 
the  bending wave radius
$r_{bw}=3(h/r)^{-4/5}a_*^{2/5}$, $\Sigma_0$ is a dimensionless normalisation
constant and $\dot{M}_0$ is the average mass accretion rate which we will
assume stays constant over the course of a single observation. This gives
$\Sigma \propto r^\lambda$ for small $r$ and $\Sigma \propto r^{-\zeta}$ for
large $r$, where $\kappa$ governs the sharpness of the break. The 
bending wave radius
occurs at radii larger than the last stable orbit because there are
additional torques created by the misaligned black hole spin
which result
in additional stresses i.e. enhanced angular momentum transport. The
material in falls faster, so its surface density drops.

Mass conservation then sets the viscous frequency as
\begin{equation}
f_{visc}(r_n)= \frac{\dot{M}_0}{2\pi R^2 \overline{\Sigma(r,t)}}
= \frac{1}{2\pi r_{bw}^2 \Sigma_0}
\frac{(1+x^\kappa)^{(\zeta+\lambda)/\kappa}}{x^{\lambda+2}}\frac{c}{R_g},
\label{eqn:fvisc}
\end{equation}
such that $f_{visc}\propto r^{\zeta-2}$ for large $r$ and
$f_{visc}\propto r^{-(\lambda+2)}$ for small $r$.

\subsection{Propagating mass accretion rate fluctuations}
\label{sec:mdot}

As in ID11 (and AU06), we start by splitting the flow up into $N$
annuli of width $dr_n$ such that $r_1=r_o$ (the truncation radius) and
$r_N = r_i+dr_n \approx r_i$ (the inner radius of the flow). We assume
that the power spectrum of variability generated in mass accretion
rate at the $n^{th}$ annulus is given by a zero centred Lorentzian
cutting off at the local viscous frequency
\begin{equation}
|\tilde{\dot{m}}(r_n,f)|^2 \propto {1\over 1+ 
(f/f_{visc}(r_n))^2 }
\label{eqn:poff}
\end{equation}
where a tilde denotes a Fourier transform and 
$f_{visc}(r_n)$ is derived from Equation \ref{eqn:fvisc}. 

We use the method of Timmer \& Koenig (1995) to generate mass
accretion rate fluctuations, $\dot{m}(r_n,t)$, which satisfy equation
\ref{eqn:poff}. These are normalised to have a mean of unity and
fractional variability $\sigma/I = F_{var}/\sqrt{N_{dec}}$ where,
unlike ID11, $F_{var}$ and $N_{dec}$ are the fractional variability
and number of annuli per decade in {\sl viscous frequency} rather than
radius. These two descriptions are exactly equivalent where $f_{visc}$
is a power law function of radius as in ID11, as $df/f=dr/r$. However,
the more physical smoothly broken power law form for $f_{visc}$ does
not retain this property. We choose to parametrise the noise power in
terms of $df/f$ (see below). 

The mass accretion rate through the outer annulus is given by
$\dot{M}(r_1,t)=\dot{M}_0\dot{m}(r_1,t)$. Variability is generated in every
other annulus according to Equation \ref{eqn:poff}, but this is
also accompanied by the 
noise from the outer regions of the flow which propagates
inwards. Thus the mass accretion rate at the $n^{th}$ annulus is
given by
\begin{equation}
\dot{M}(r_{n},t)=\dot{M}(r_{n-1},t-t_{lag}) \dot{m}(r_{n},t),
\end{equation}
where $t_{lag}=-R_g dr_n/v_r(r_n)=dr_n/(r_nf_{visc}(r_n))$ is the
propagation time across the $n^{th}$ annulus and $v_r(r_n)=-R_g r_n
f_{visc}(r_n)$ is the infall velocity.

To convert these mass accretion rate fluctuations into a lightcurve, 
we assume that the 
luminosity emitted from the $n^{th}$ annulus is given by
\begin{equation}
dL(r_n,t) = \eta/2 ~ \dot{M}(r_n,t) c^2 \epsilon(r_n) r_n dr_n,
\end{equation}
where the (dimensionless) emissivity is given by
\begin{equation}
\epsilon(r_n)=\epsilon_0 r_n^{-\gamma}b(r_n),
\end{equation}
and $\gamma$ is the emissivity index, $\eta$ the accretion efficiency,
$b(r)$ the boundary condition and $\epsilon_0$ is a normalisation
constant. 
In ID11, we considered two boundary
conditions: the `stress free' boundary condition
$b(r)=3(1-\sqrt{r_n/r_i})$ and the `stressed' boundary condition
$b(r)=1$. Here, however, we make the intuitive and physical assumption
that the boundary condition is set by the surface density such that
$b(r)\propto \overline{\Sigma(r_n,t)}$ where
$\overline{\Sigma(r_n,t)}$ is the \textit{time averaged} surface
density. This allows the model to link the emission with the amount of
material in a particular annulus.

The fluctuating mass accretion rate will also have an effect on precession
because mass conservation needs to hold on short time scales as well as
long time scales, which gives $\dot{M}(r_n,t) \propto \Sigma(r_n,t)2\pi r^2
f_{visc}$. This means that the surface density at time $t$ is given by
\begin{equation}
\Sigma(r_n,t)= \frac{\Sigma_0 \dot{M}(r_n,t)}{cR_g}
\frac{x^\lambda}{(1+x^\kappa)^{(\zeta+\lambda)/\kappa}},
\label{eqn:sigma}
\end{equation}
which trivially averages to equation \ref{eqn:sigmabar} on long time scales.
Because the surface density sets the precession frequency (equation
\ref{eqn:fprec}), we see that the fluctuations in mass accretion rate cause
the precession frequency to vary, thus allowing the model to predict a
quasi-periodic oscillation rather than a purely periodic oscillation.

\begin{figure}
\centering$
\begin{array}{c}
\leavevmode  \epsfxsize=6.cm \epsfbox{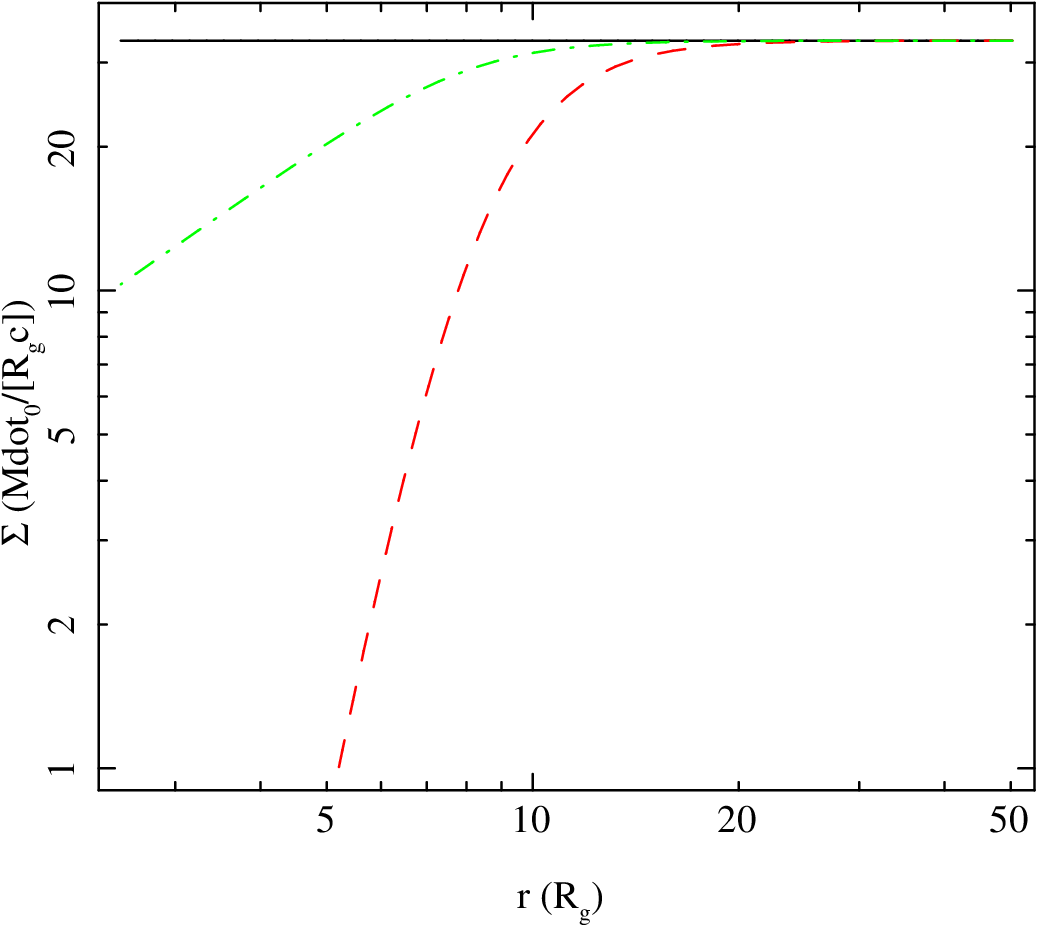}\\
\leavevmode  \epsfxsize=6.cm \epsfbox{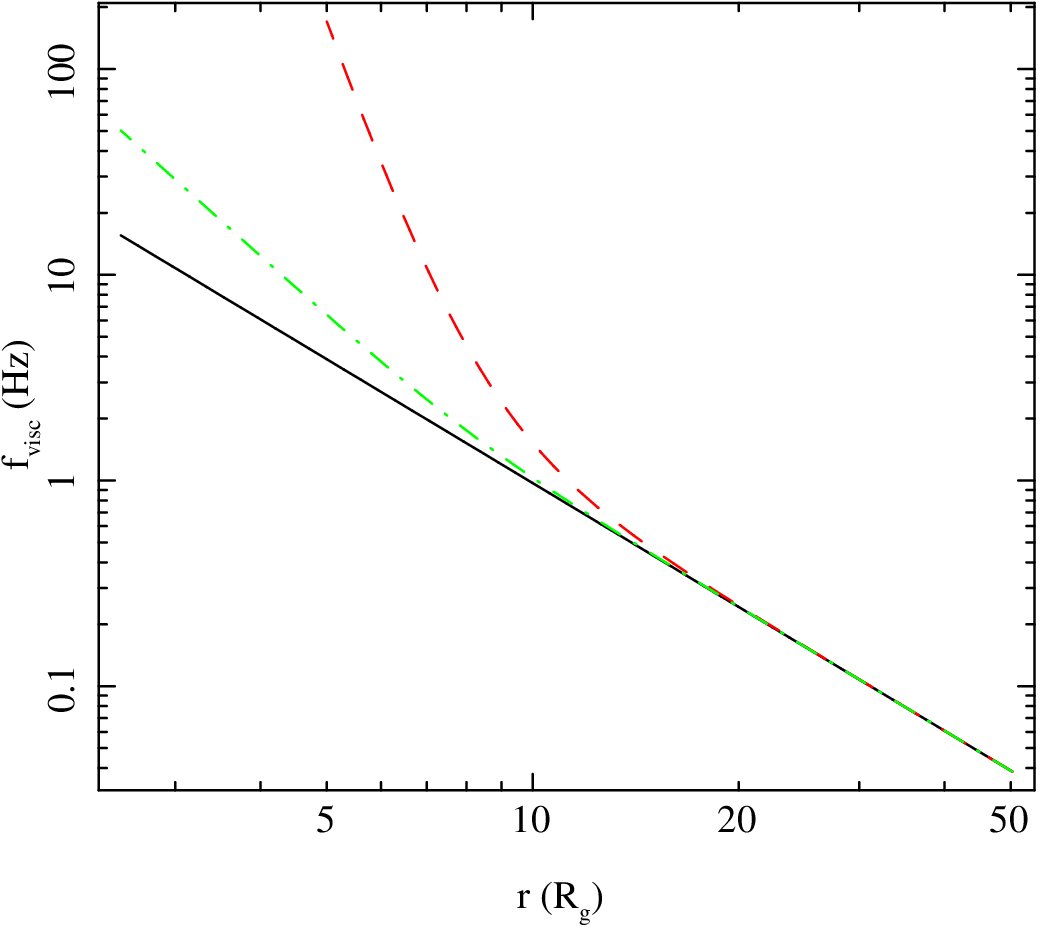}\\
\leavevmode  \epsfxsize=6.cm \epsfbox{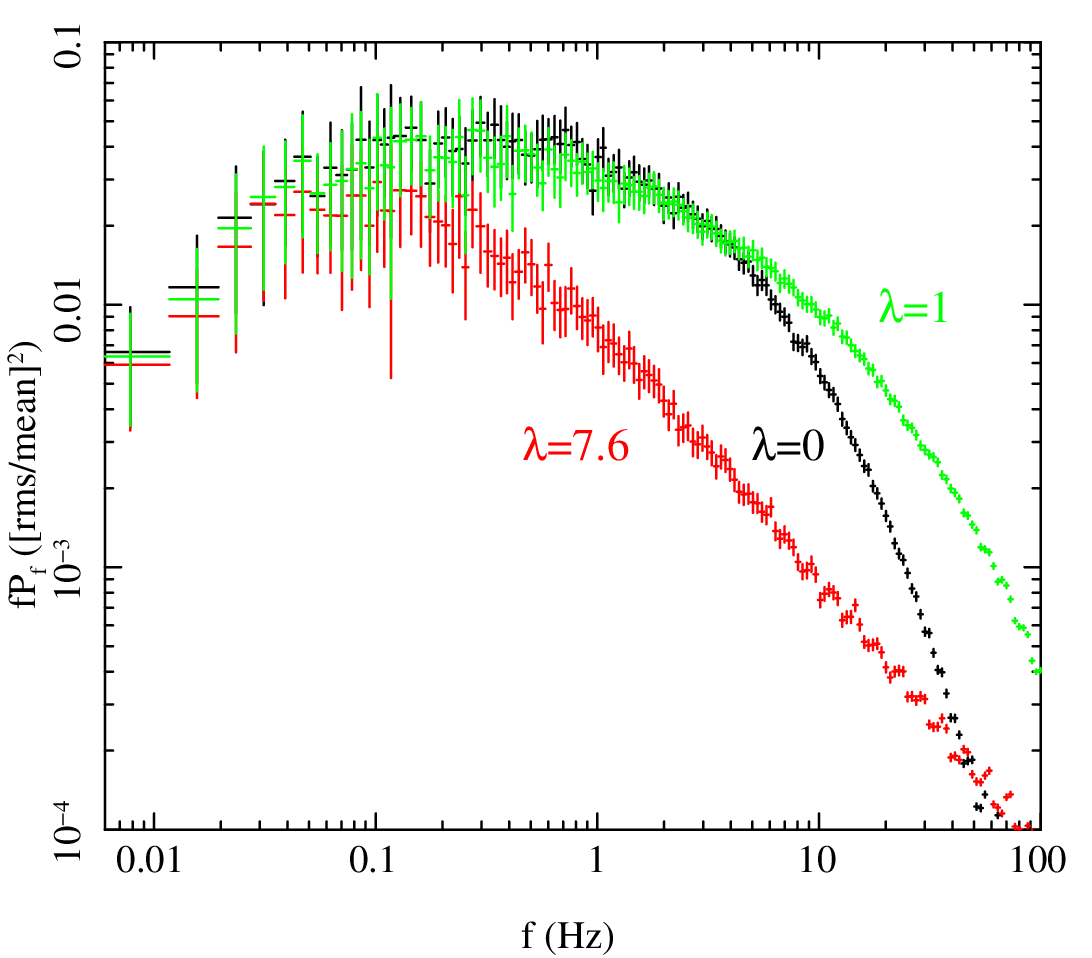}
\end{array}$
\caption{
\textit{Top (a):} Surface density as a function of radius for the
fiducial model parameters in ID11 (solid black line), simulations
of a misaligned accretion flow around a $10$ solar mass black hole
with $a_*=0.5$ (red dashed line) and the fiducial model parameters
we choose for this paper (green dot-dashed line). The red dashed line
is calculated using equation \ref{eqn:sigmabar} with $\lambda=7.6$, $\kappa=5$,
$\zeta=0$ and $r_{bw}=8.08$ (the parameters which best fit the simulation data).
For the dot-dashed green line, $\lambda=1$ with all other parameters the same.
\textit{Middle (b):} The viscous frequency as a function of radius resulting
from assuming the surface density to be given by the corresponding line in the
top panel.
\textit{Bottom (c):} The PSD predicted using the surface density given by
corresponding lines in the top panel.}
\label{fig:sigmap}
\end{figure}

\subsection{Surface density profile}

In ID11, we parametrised the viscous frequency with a power
law. The fiducial model parameters therein gave $f_{visc}=0.03
r^{-0.5} f_k$, corresponding to a surface density profile
$\Sigma(r)\propto r^{0}$ between the inner and outer radii which were
set to $r_i=2.5$ and $r_o=50$ respectively.  By comparison, the GRMHD
titled flow simulations of Fragile et al (2009) also give
$\Sigma(r)\propto r^{0}$ at large radii, but then smoothly break at
the bending wave radius to a much steeper dependence. The most
relevent simulation to this paper is the case with $a_*=0.5$ as this
is likely closest to the spin of XTE J1550-564 (e.g. Davis, Done \&
Blaes 2006; Steiner et al 2011).
This has surface density parameters (Equation
\ref{eqn:sigmabar}, see Figure 4 in IDF09) of $r_{bw}=8.1$
(corresponding to $h/r=0.21$), $\kappa=5$, $\lambda=7.6$ and
$\zeta=0$.

In Figure \ref{fig:sigmap} a and b, we plot these  two different surface
density prescriptions and their resulting viscous frequencies, with
the power law shown by the black solid line and the broken power law
shown by the red dashed line. In the case of the broken power law, we
choose the normalisation $\Sigma_0=33.3$ to ensure that both assumptions
become consistent with one another at large radii. Figure \ref{fig:sigmap}
c shows the PSD resulting from the two different prescriptions. The new
(and more physically realistic) surface density prescription predicts much
less  noise at high frequencies than the previous model, where the 
surface density remained constant down to the innermost radii. 
This distinction is even more striking as our assumption that the MRI
noise power is constant per decade in frequency (as opposed to radius) 
means that the regions with $r<<r_{bw}$ have more MRI power generated
per decade (in radial extent) than those with $r>>r_{bw}$. Yet even this
additional power at small radii is not sufficient to give enough high frequency
power to match that seen in the data. 

To retrieve sufficient high frequency power requires $\lambda=1$
rather than $7.6$ (green dot-dashed lines in Fig \ref{fig:sigmap} a
and b). This gives a more gradual drop-off in surface density, leading
to a less severe transition in viscous frequency at the bending wave
radius and hence more high frequency power (green points in Fig
\ref{fig:sigmap}c).  We discuss the physical implications of this in
more detail in Section \ref{sec:disc}. For now, however, we use
$\lambda=1$ for our fiducial model.

\begin{figure}
\centering$
\begin{array}{c}
\leavevmode  \epsfxsize=6.cm \epsfbox{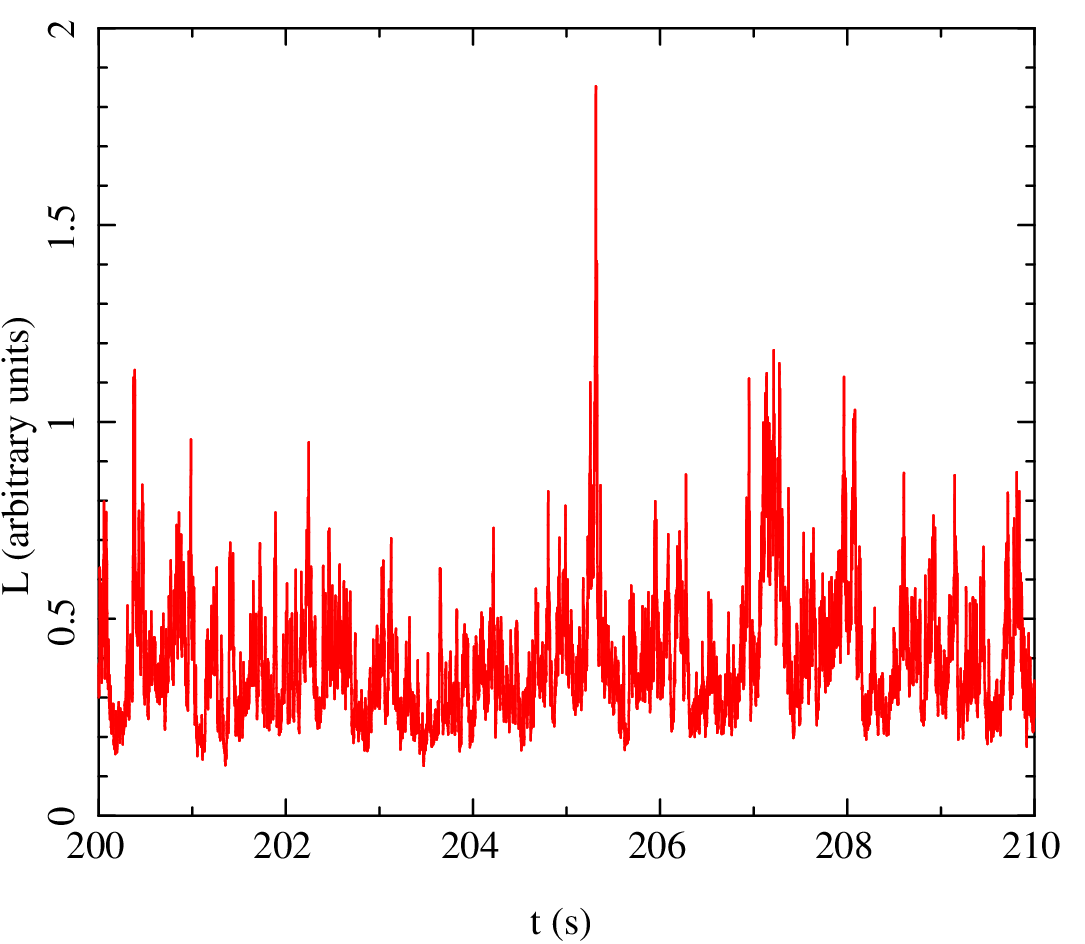}\\
\leavevmode  \epsfxsize=6.cm \epsfbox{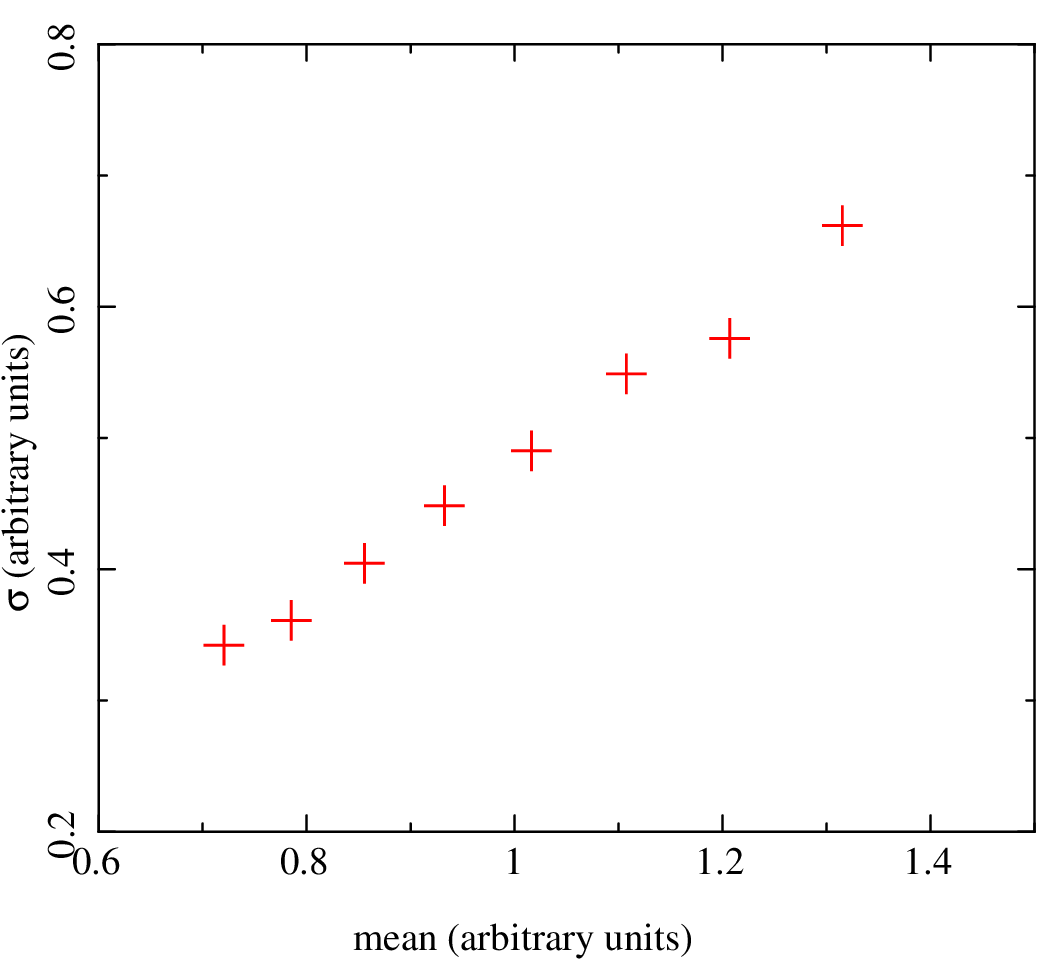}
\end{array}$
\caption{
\textit{Top (a):} A 10s segment of the light curve calculated using the fiducial model
parameters and $r_o=50$.
\textit{Bottom (b):} The sigma-flux relation for the above light curve. We see this
is linear as is seen in the data.}
\label{fig:lc}
\end{figure}

\begin{figure}
\centering
\leavevmode  \epsfxsize=6.5cm \epsfbox{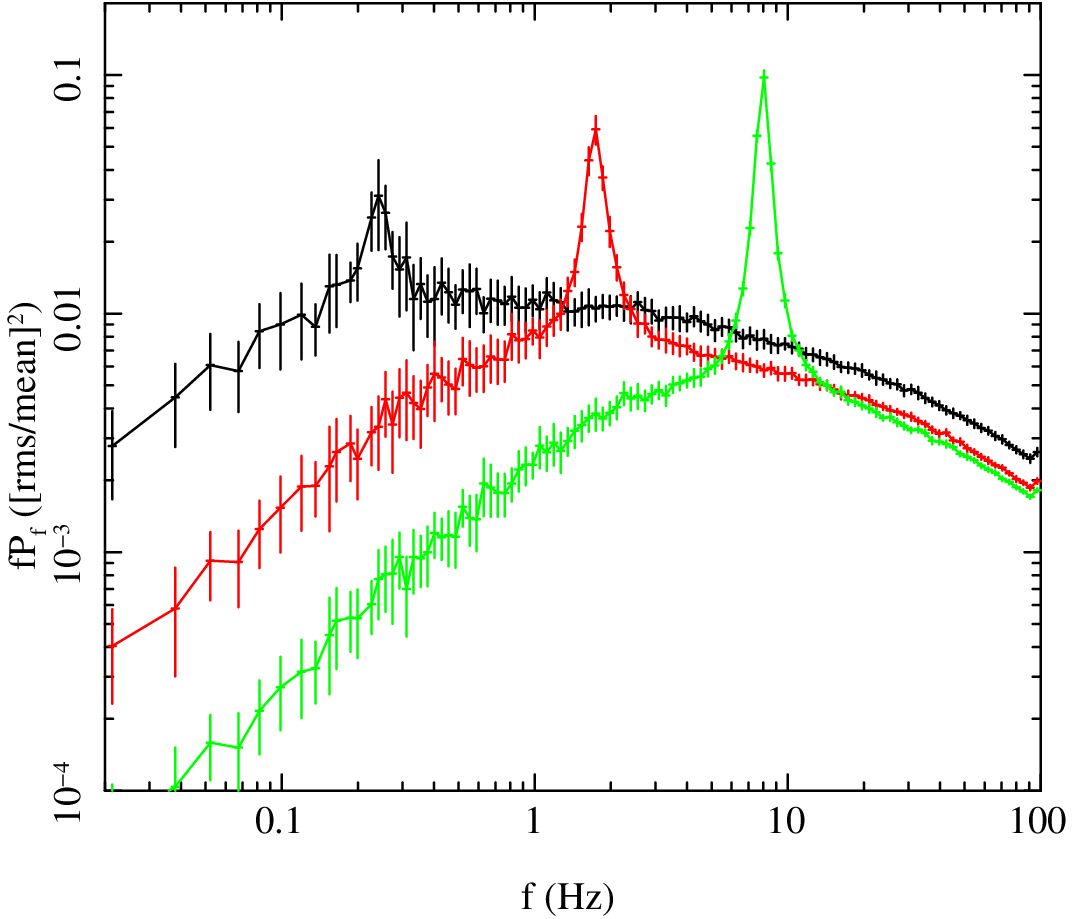}
\caption{The predicted PSD for the fiducial model parameters with $r_o=50$ (black),
$20$ (red) and $10$ (green). For clarity the normalisation of the fundamental is set
to increase as $r_o$ reduces and all the other QPO components are normalised to zero.}
\label{fig:3power}
\end{figure}

\subsection{The fiducial model}
\label{sec:fiducial}

Following the discussion in section \ref{sec:sigma}, we use model parameters
$\Sigma_0=33.3$, $r_{bw}=8.1$, $\kappa=5$, $\lambda=1$ and $\zeta=0$. We also
set $r_i=2$ and $\gamma=4$ but note that the new assumptions for surface
density coupled with the new boundary condition mean the model is now
much less sensitive to the parameter $r_i$ than its predecessor in ID11.
Figure \ref{fig:lc}a shows a $10s$ segment of the light curve created
using these assumptions and with $r_o=50$. We use $N_{dec}=15$ (i.e. $15$
annuli per decade in viscous frequency) with $2^{22}$ time steps, giving a
duration of $4096s$ (similar to a typical RXTE observation) for a time bin
of $dt=9.7\times 10 ^{-4}$. Figure \ref{fig:lc}b confirms that this light
curve has the linear sigma-flux relation implied by its skewed nature. The
PSD of this light curve is represented by the red points in Figure
\ref{fig:sigmap}c.

We calculate the QPO as in ID11, but we briefly summarise this here
for completeness. The QPO fundamental frequency is set to the average
(over the $4096s$ duration) precession frequency calculated from
Equation \ref{eqn:fprec}. In principle we can calculate the width of
the QPO from the fluctuations in frequency which result from
fluctuations in surface density. However, these only set a lower limit
to the width of the QPO since it can also be broadened by other
processes (ID11), so we leave this as a free parameter.  We can in
principle predict the harmonic structure in the QPO lightcurve by a
full Comptonisation calculation of the angle dependent emission from a
precessing hot flow (Ingram, Done \& Zycki in preparation). Until
then, we simply allow the normalisations of the harmonics to be a free
parameter but fix their width so that they have the same quality
factor as the fundamental (apart from the sub-harmonic which is free:
Rao et al 2010). We use the method of Timmer \& Koenig (1995) to generate a
light curve from these narrow QPO Lorentzians and add this to the
light curve already created for the broad band noise.

Figure \ref{fig:3power} shows the full PSD given by the fiducial model
parameters with $r_o=50$ (black), $r_o=20$ (red) and $r_o=10$
(green). For clarity we set the normalisations of the QPO harmonics to
zero, and increase the normalisation and quality factor of the
fundamental as $r_o$ decreases to match with the data. This captures
the essence of the observed evolution of the PSD in terms of a
decreasing truncation radius.

\section{Fitting to data}
\label{sec:data}

\begin{figure}
\centering$
\begin{array}{c}
\leavevmode  \epsfxsize=5.5cm \epsfbox{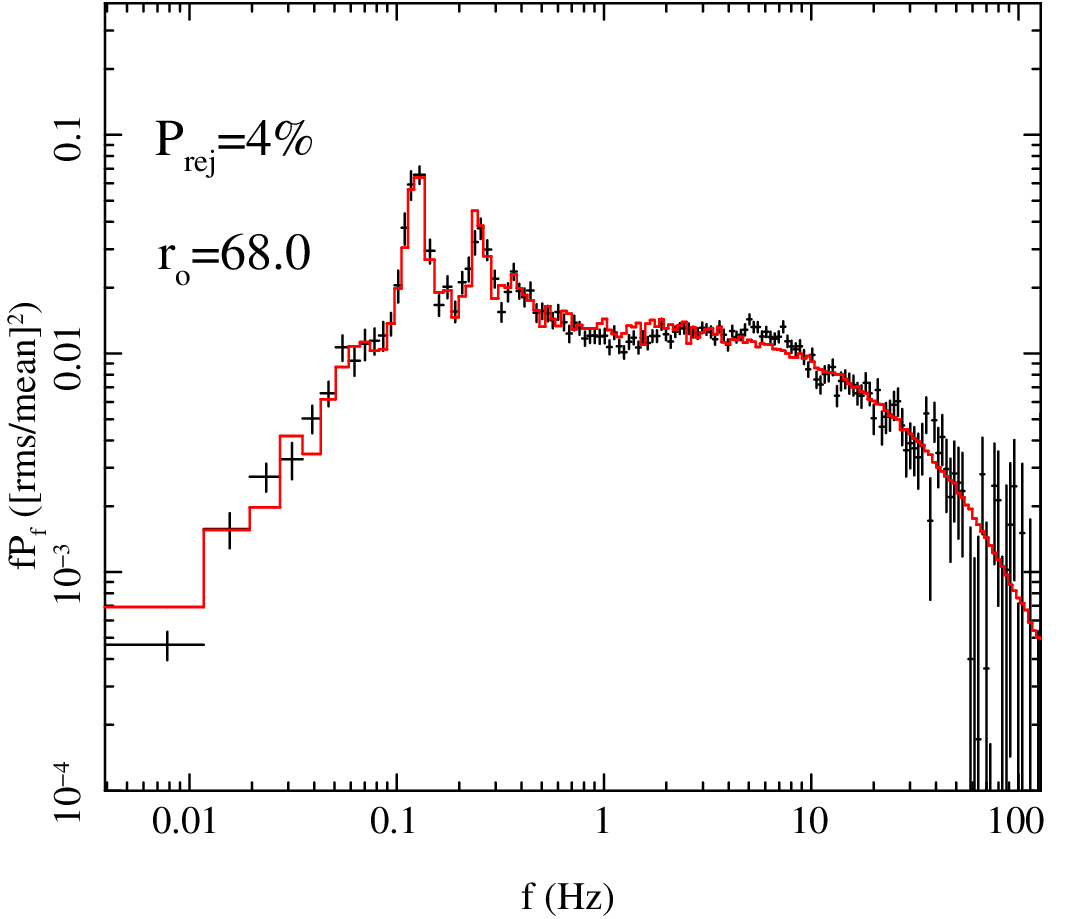}\\
\leavevmode  \epsfxsize=5.5cm \epsfbox{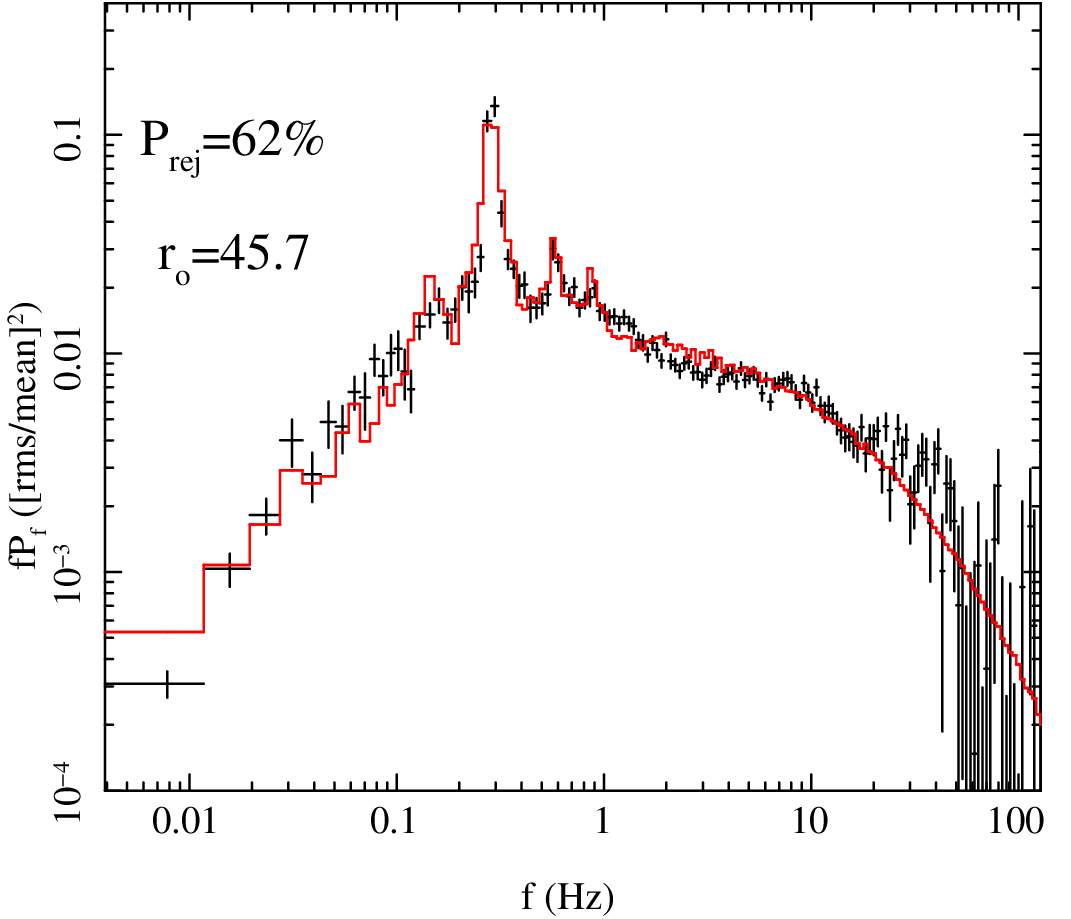}\\
\leavevmode  \epsfxsize=5.5cm \epsfbox{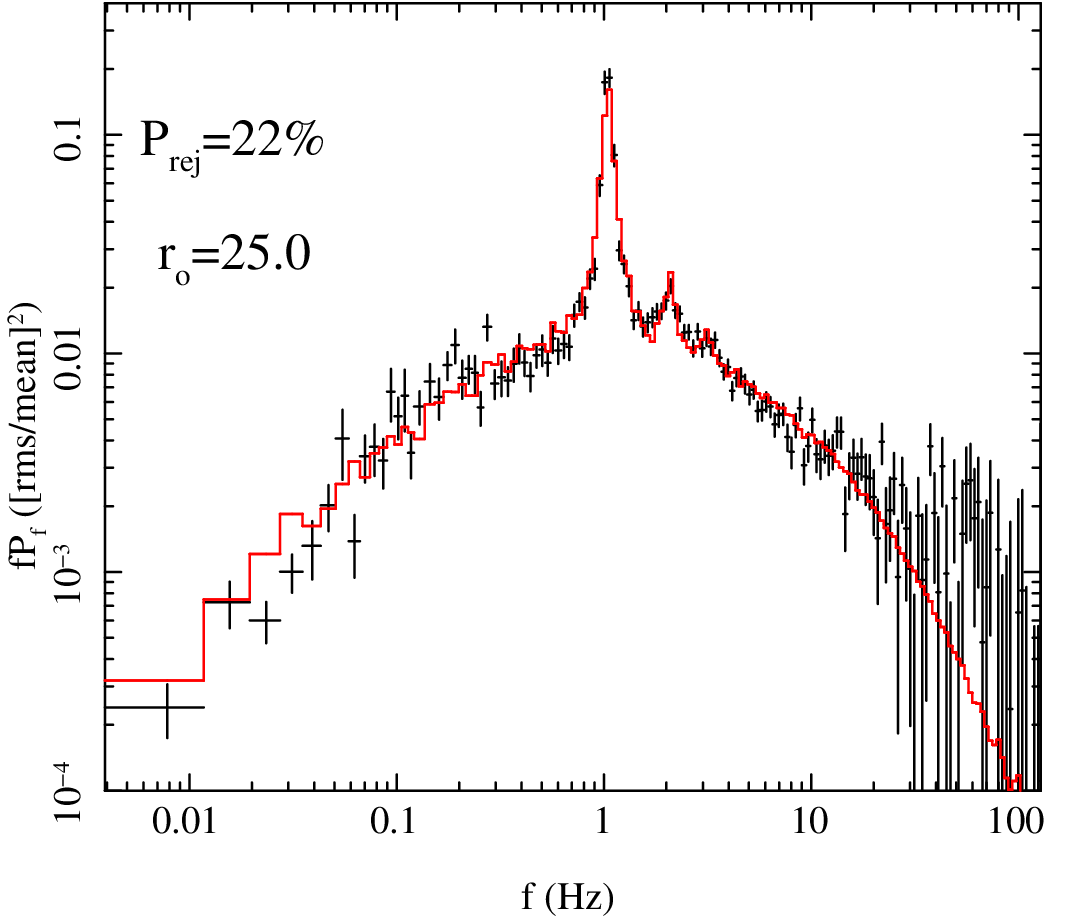} \\
\leavevmode  \epsfxsize=5.5cm \epsfbox{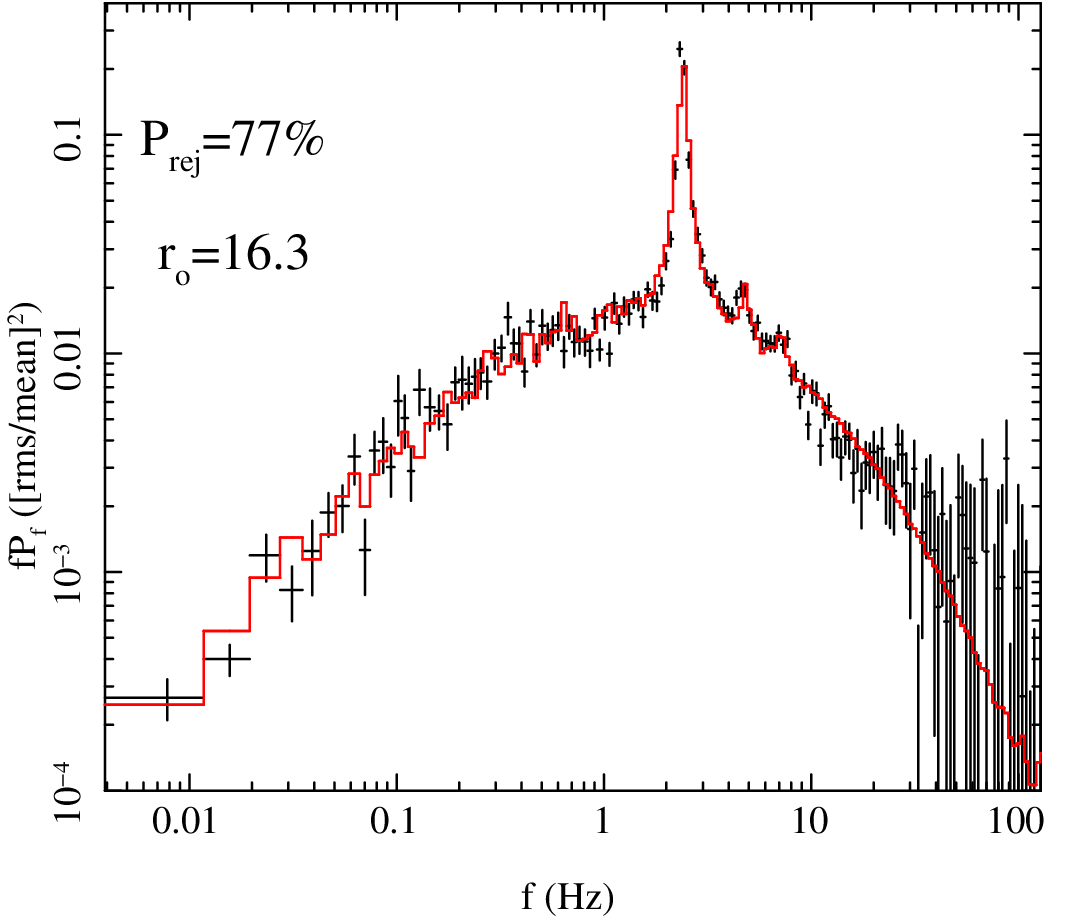} \\
\leavevmode  \epsfxsize=5.5cm \epsfbox{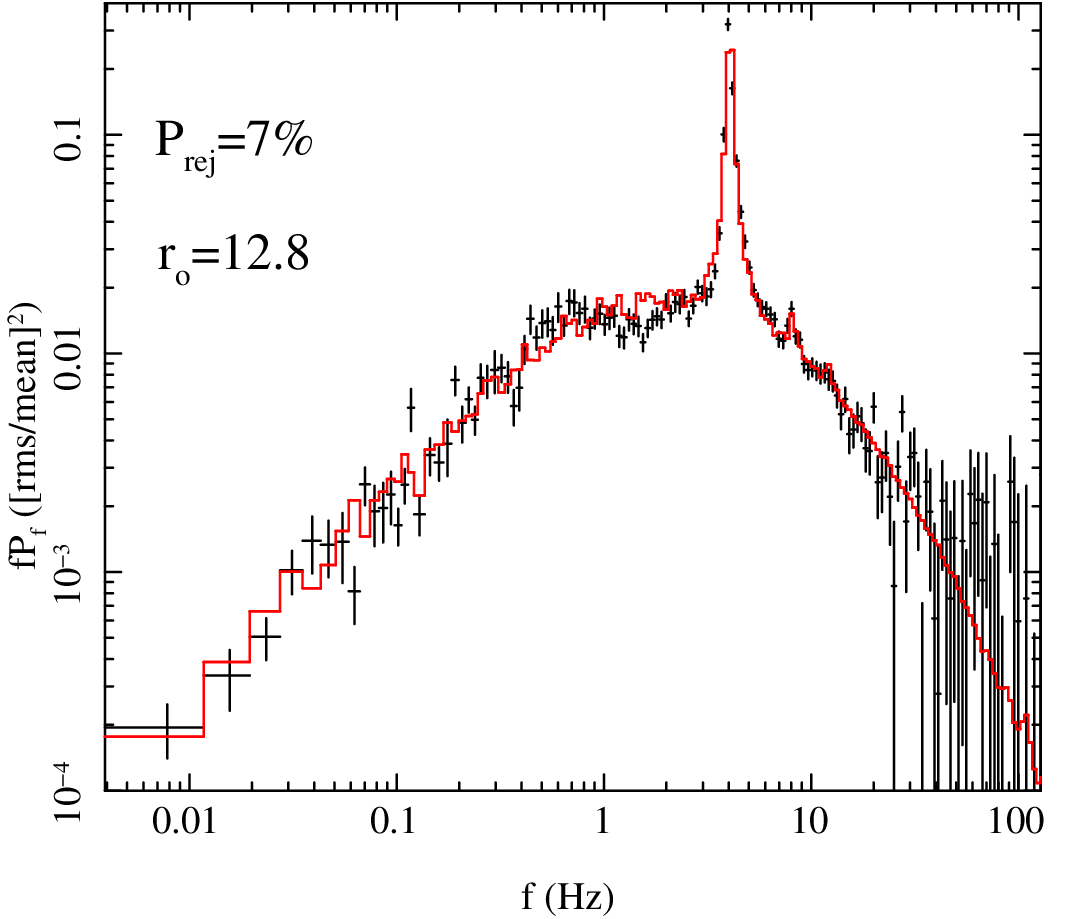}
\end{array}$
\caption{Best fit PSDs along with data points for observations 1-5 (top
to bottom respectively). The rejection probability, $P_{rej}$, and
truncation radius, $r_o$, are included in each plot. The rest of the
best fit physical parameters are included in table \ref{tab:results}.}
\label{fig:fits}
\end{figure}

The main problem with power spectral fitting is that we can only
calculate the power spectrum of an infinitely long light curve which,
of course can neither be observed nor simulated! The discreet version
of the power spectrum is the periodogram which provides an estimate
of the power spectrum, but a very poor one with large, non-Gaussian
errors. There are two main techniques for smoothing the periodogram
in order to improve the statistics (i.e. smaller and more Gaussian errors).
Firstly, the lightcurve can be split up into shorter segments with the
PSD estimate given by averaging the periodogram over all segments.
Secondly (or additionally) the linearly sampled PSD can be averaged into
logarithmic/geometrically spaced frequency intervals (e.g. van der Klis 1989).

In ID11, we averaged the logarithm of the periodogram in order to get
a PSD estimate which, for AGN or any red noise power spectra, gives an
unbiased estimate with errors close to Gaussian for a small amount of
smoothing compared with averaging the periodogram linearly (Papadakis
\& Lawrence 1993). However, the BHB power spectra are better described
by band limited noise, with high $d^2P(f)/df^2$. In this case,
averaging the logarithmic periodogram gives a biased estimator
(because the bias is proportional to $d^2P(f)/df^2$; Papadakis \&
Lawrence 1993). Because of this, we experienced some statistical
difficulties, the most obvious being apparent over-fitting with
reduced $\chi^2$ falling well below unity (indicating a good fit) for
parameters which have a high rejection probability.  Here instead we
average the periodograms linearly and test to see how reliable
$\chi^2$ is as a measure of goodness of fit in this case. This also
has the added advantage that we can compare to the white noise
subtracted PSD from the data, whereas, because logarithmic power must
be positive definite, we previously had to add white noise to the
model.

We split each simulated model lightcurve of $4096$s into $32$ segments
each of duration $128$s, so the lowest frequency we can resolve is
$1/128~$Hz. We average these at each linearly spaced frequency bin,
$j$ to get $P_{mod}(f_j)$ and the dispersion around this mean $\Delta
P_{mod} (f_j)$. We then re-bin the result in the same way as the data
in geometrically separated frequency bins, $f_J$, so that the model
power is $P_{mod}(f_J) = \Sigma P_{mod}(f_j)/\Delta P^2_{mod}(f_j)$
and its propagated error is $\Delta P_{mod}(f_J) = \Sigma 1/\Delta
P^2_{mod}(f_j)$ where the sums are taken over all the $f_j$ which fall
inside the bin width of $f_J$.  The lowest geometric
frequency bin typically only contains one linear frequency bin, so is
an average over only 32 points. While this would not be Gaussianly
distributed for red noise (Papadakis \& Lawrence 1993) our noise power
spectra at these low frequencies are typically white, so this is
approximately Gaussian. At higher frequencies, the PSD is made from
averaging many more estimates, so is again approximately
Gaussian. Thus we should get a good estimate for the best fit by
minimising $\chi^2$ between the model and data using
\begin{equation}
\chi^2 = \sum_J \frac{(P_{mod}(f_J)-P_{obs}(f_J))^2}
{\Delta P^2_{mod}(f_J)+\Delta P^2_{obs}(f_J)}
\end{equation}
where $P_{obs}(f_J)$ is the PSD of the data, and $\Delta P_{obs}(f_J)$
is the propagated error from the (linear) averaging of the multiple
segments and geometric re binning. For the data, the number of
segments dependends on the observation length rather than being fixed
as in the model. For our lightcurves this gives $41$, $26$, $13$, $14$
and $14$ segments for observations $1-5$ respectively. The rather
small number of segments in observations 3-5 mean that the errors on the
data at the lowest frequencies may only be approximately Gaussian, but the
total error is given by the sum of those from {\em both} the model and
the data, so the fit statistic should remain close to a $\chi^2$
distribution, so minimising $\chi^2$ should indeed return the best fit. 

We have coded the entire model into {\sc xspec} for public
release as {\sc propfluc}, described in detail in the Appendix.

\section{Example fits to XTE~J1550-564}
\label{sec:data}

We use the same data as in ID11 so that we can directly compare
results i.e. \textsc{rxte} data from the 1998 rise to outburst of XTE
J1550-564 (Sobczak et al 2000; Wilson \& Done 2001; Remillard et al
2002; Rao et al 2010; Altamirano 2008) from Obs IDs: 30188-06-03-00,
30188-06-01-00, 30188-06-01-03, 30188-06-05-00 and 30188-06-11-00
(hereafter observations 1-5 respectively). We consider energy channels
36-71 (i.e. 10-20keV) in order to avoid any direct
contamination from the disc emission. 

We fit each observed PSD to derive the parameters of the smoothly
broken power law surface density. We assume that the shape of the
surface density stays constant across all datasets, but its
normalisation $\Sigma_0$ can change. We also allow the bending wave
radius to be a free parameter, $r_{bw}=3(h/r)^{-4/5}a_*^{2/5}$ (where
$h/r$ is the scaleheight of the flow). As we have fixed the spin, the
best fit value of $r_{bw}$ gives us an estimate of the scale height of
the flow which may change through the transition due to the increase
in seed photons from the disc cooling the flow.  The inner radius of
the flow is tied across all the data sets, and we fit for $r_o$.  The
remaining free parameters which determine the broadband noise are the
level of MRI fluctuations generated over each decade in frequency,
$F_{var}$, and the emissivity index, $\gamma$ (held constant across
all 5 observations).

While {\sc xspec} can fit the model to the 5 PSD simultaneously, this
is very slow. Instead, we used trial and error to set values of the 
parameters which are tied across all the datasets and then fix these
to fit the remaining parameters for each PSD individually.

\subsection{Fit results}
\label{sec:results}

\begin{table*}
\begin{tabular}{l|l|l|l|l|l|l|l|l|l|c}
 \hline
  Obs & $\Sigma_0 $& $\zeta$  & $\lambda$ & $\kappa$ & $r_i$ & $r_o$ & $h/r~(r_{bw})$ & $F_{var}$ & $\gamma$

 \\
 \hline
 \hline
1  & $5.43$   &$ $         & $    $        & $  $          & $  $          & $68.0$ & $0.41~(4.6)$   & $0.32$ & $    $ \\

2  & $10.48$  &$ $         & $    $        & $  $          & $  $          & $45.7$ & $0.27~(6.5)  $ & $0.31$ & $    $ \\

3  & $21.73$  &$\equiv 0$  & $\equiv 0.9$  & $\equiv 3.0$  & $\equiv 3.3 $ & $25.0$ & $0.21~(8.0)$   & $0.36$ & $5.28$ \\

4  & $30.03$   &$ $         & $    $        & $  $          & $  $          & $16.3$ & $0.13~(12.03) $ & $0.43$ & $    $ \\

5  & $30.36$  &$ $         & $    $        & $  $          & $  $          & $12.8$ & $0.12~(12.1) $ & $0.48$ & $    $ \\

\hline
\end{tabular}
\caption{Best fit physical parameters for observations 1-5. A $\equiv$ symbol indicates
that the parameter has been fixed.}

\label{tab:results}
\end{table*}     

\begin{figure}
\centering
\leavevmode  \epsfxsize=6.5cm \epsfbox{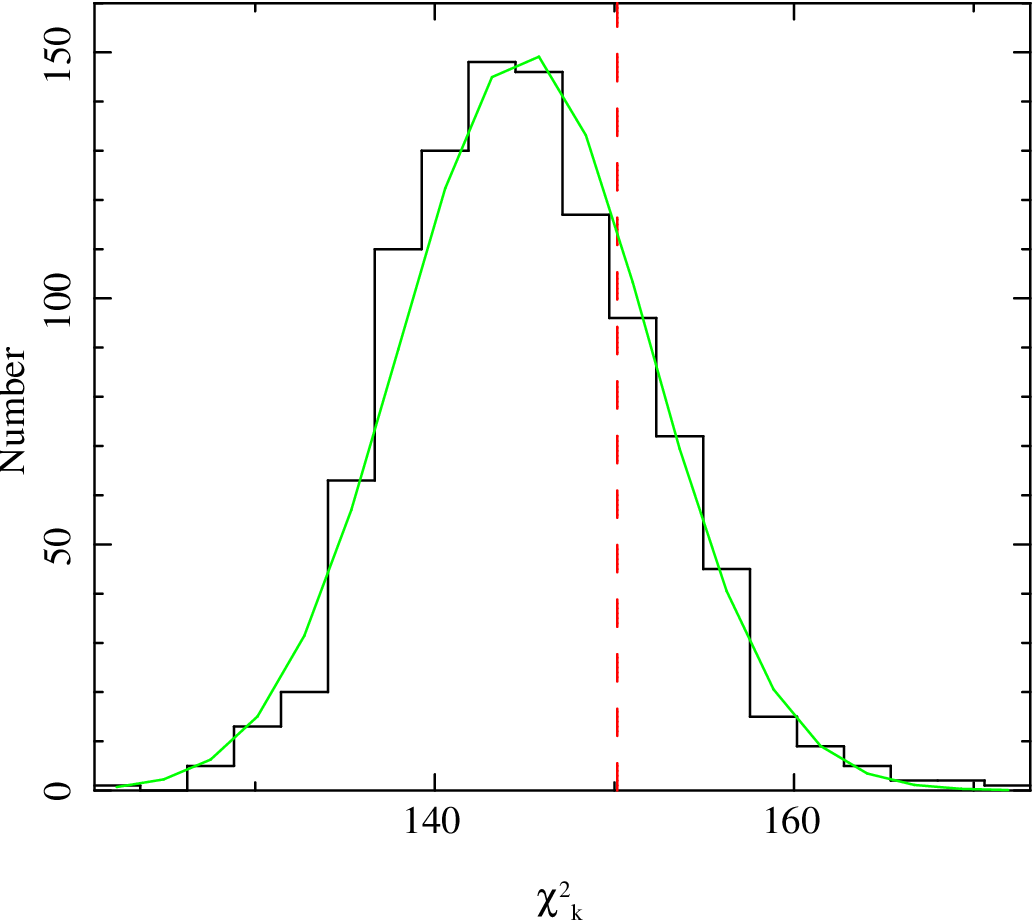}
\caption{Distribution of $\chi^2_k$ values calculated using the
best fit model parameters for observation 4. The green line
illustrates that this is a nearly Gaussian distribution. The
red dashed line picks out the $\chi^2$ value for this observation
and we see that, although it is larger than the mean $\chi^2_k$
value, it still lies believably within the distribution meaning
we can be confident that the model fits. }
\label{fig:hist4}
\end{figure}

The data and best fit model PSD are shown in 
Figure \ref{fig:fits}. These
give a  reasonable reduced $\chi^2$ 
value of $1.09$ (764.6 for 704 degrees of freedom), unlike ID11. 
We check the goodness of fit {\it a posteriori}
by calculating the rejection probability (Uttley et al 2002; Markowicz
et al 2003). We do this by taking the minimised $\chi^2$ value
\begin{equation}
\chi^2 = \sum_f \frac{(P_{mod}(f)-P_{obs}(f))^2}
{\Delta P_{mod}^2+\Delta P_{obs}^2}
\end{equation}
where $P_{obs}(f)$ is the PSD estimate for the observed data and
$P_{mod}(f)$ is the PSD estimate for one particular realisation
of the model. We then simulate many more (1000) realisations with
the same model parameters in order to calculate many values of
\begin{equation}
\chi^2_k = \sum_f \frac{(P_{mod}(f)-P_{k}(f))^2}
{\Delta P_{mod}^2+\Delta P_{k}^2}
\end{equation}
where $P_{k}(f)$ is the PSD estimate for the $k^{th}$ realisation.
The rejection probability, $P_{rej}$, is given by the percentile of
$\chi^2_k$ values which are smaller than $\chi^2$. This does not
assume that the errors are Gaussian, so is more strictly accurate as
it assesses the likelihood that $P_{obs}(f)$ does not belong to the
distribution which $P_{mod}(f)$ and each $P_{k}(f)$ belong to. We
derive $P_{rej}=4\%$, $62\%$, $22\%$, $77\%$ and $7\%$ for
observations 1-5 respectively. The lowest values of $P_{rej}$
obviously imply a very good fit but even the higher values are still
acceptable. Figure \ref{fig:hist4} shows the distribution of
$\chi^2_k$ values from the $P_{rej}$ calculation (black stepped line)
using the best fit parameters for observation 4. The red dashed line
shows the $\chi^2$ value for this observation and we see that,
although it is larger than most $\chi^2_k$ values, it still lies
believably within the distribution.  We also plot (green solid line) a
Gaussian with the same mean, standard deviation and normalisation as
the distribution and we see very good agreement between the two. This
confirms that the PSD estimate we use does indeed give (approximately)
Gaussian errors and therefore $\chi^2$ is a reliable measure of
goodness of fit.

Table \ref{tab:results} shows all of the best fit physical
parameters. Some of these parameters are very similar to those derived
from the previous model fits in ID11 e.g. the truncation radius moves
from $r_o=68-13$, while $F_{var}$ increases throughout the transition.
However, our new parametrisation means that we can directly explore
the change in bending wave radius, $r_{bw}$, and normalised surface
density $\Sigma_0$. The bending wave radius increases, implying that
the flow scaleheight, $h/r$, is collapsing.  This makes sense
physically as the decreasing truncation radius means that the flow is
cooled by an increasing number of seed photons, so the electron
temperature decreases. The spectra also show that the optical depth
increases (as is also implied by the increasing normalised surface
density). This increases the coupling between electrons and ions so
the ion temperature also decreases (Malzac \& Belmont 2009).  The flow
is held up (at least partly) by ion pressure so the scale height of
the flow collapses.

\section{Discussion}
\label{sec:disc}

\begin{figure}
\centering
\leavevmode  \epsfxsize=6.5cm \epsfbox{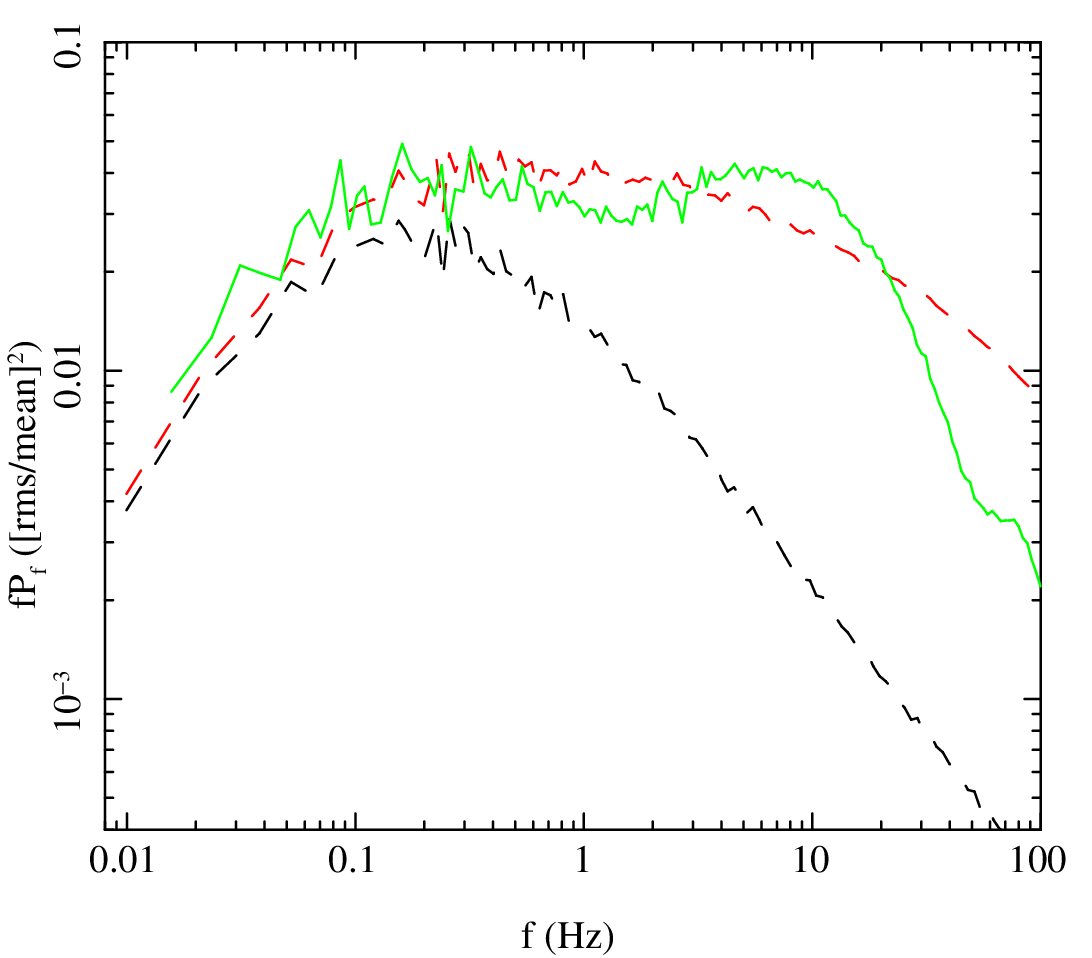}
\caption{The red dashed line is the PSD predicted using the fiducial
model parameters with $r_o=50$ (i.e. $\lambda=1$) whereas the black dashed
line is for $\lambda=7.6$ with all other parameters the same. For the
green solid line, we use the same parameters as used for the black line
but we have changed the model by assuming the annulus containing $r_{bw}$
to me more variable than the other annuli. We see we can recover the
amount of high frequency power required to match the observations using this
assumption.}
\label{fig:changemri}
\end{figure}

We have improved upon the model of ID11 by including a surface density
profile which has the same shape as predicted by GRMHD simulations. We
obtain an excellent fit to data for five observations and the
evolution of the best fit parameters is self-consistent.  However, we
require the surface density interior to the bending wave radius to
drop-off as $r^\lambda$ with $\lambda \sim 1$, where as the
simulations predict $\lambda \approx 7$ (see Figure
\ref{fig:sigmap}). The most likely reason for this apparent
discrepancy is that the torque created by the misalignment between
flow and black hole angular momenta not only creates a drop-off in
surface density but also generates extra turbulence which we do not
account for in our model.  Because the surface density sets the
emissivity, we can still reproduce the observations by over predicting
the surface density at small $r$ to compensate for under predicting
the intrinsic variability. In Figure \ref{fig:changemri}, we re-plot
the predicted PSD for $\lambda=7.6$ (dashed black line) and
$\lambda=1$ (dashed red line) without errors for clarity. For the
green solid line, also plotted without errors, we have changed the
model slightly. We again set $\lambda=7.6$ but now the fractional
variability in the annulus containing $r_{bw}$ is higher (by a factor
of 10) than that at all other annuli so as to approximate the
additional turbulence created by the bending waves. We see that it is
possible to qualitatively reproduce the shape of the broad band noise
using a surface density profile consistent with simulations if we
include this extra assumption.

It is interesting that the green line in Figure \ref{fig:changemri}
does not have a flat top between low and high frequency breaks as the
model generally predicts, but rather has a `bump' at $\sim 7$Hz and
another at $\sim 0.15$Hz. There are actually many observations of
bumpy power spectra such as this which cannot be well described by the
model in its current state (e.g. Axelsson et al 2006; Wilkinson \&
Uttley 2009). It therefore looks likely that the variability generated
by the MRI is not as uniform as we naively assume and actually some
regions produce more variability than others thus giving rise to a
bumpy power spectrum such as the green line in Figure
\ref{fig:changemri}. 

\section{Conclusions}
\label{sec:conc}

We have made some improvements to a model that can predict the power spectral
behaviour of BHBs in the context of the truncated disc / hot inner flow
geometry designed to explain the energy spectral evolution. The model now
assumes a surface density profile consistent with that predicted by GRMHD
simulations. This allows us to gain more physical insight from the evolution
of best fit parameters which reproduce the observed evolution of the PSD. A
coherent picture is now emerging: as the truncation radius, $r_o$, moves
inwards, the increased number of seed photons incident on the flow cool it thus
reducing both the electron and ion temperatures, $T_e$ and $T_i$ respectively.
The Comptonised emission from the flow is therefore softer and, in addition to
this, the lower ion temperature gives rise to a lower pressure meaning that the
scale height of the flow, $h/r$, should collapse. The bending wave radius, which
sets the shape of the surface density, is given by $r_{bw}=3(h/r)^{-4/5}a_*^{2/5}$
and therefore increases as $h/r$ collapses. Also, because the volume of the flow
is reducing, the surface density must increase and, by mass conservation, the
infall velocity decreases. When we fit the model to five observations of XTE
J1550-564, we see all of these trends: $r_o$ reduces and $r_{bw}$ increases as
does $\Sigma_0$, the normalisation of the surface density (and also the inverse
of the normalisation of the infall velocity).

It is also worth re-iterating some other successes of the model first addressed in
ID11. The truncated disc geometry fundamentally predicts an inhomogeneous electron
temperature across the radial extent of the flow and therefore the Comptonized
emission from the flow should actually be inhomogeneous also. Specifically, emission
from large radii should be softer than that from small radii because the outer
regions of the flow can see more seed photons than the inner regions. This is quite
a subtle effect and, due to the degenerate nature of spectral fitting, it is only
now becoming possible to constrain inhomogeneous emission from the SED alone (Makishima
et al 2008; Takahashi et al 2008; Kawabata \& Mineshige 2010). However, it is has
long been possible to observe this effect using the technique of frequency resolved
spectroscopy, which involves constraining the SED for a given temporal frequency range.
Revnivtsev, Gilfanov \& Churazov (1999) show that the SED of the fast variability is
clearly harder than that of the slow variability. This makes sense in the truncated
disc geometry because the fast variability predominantly comes from the inner regions
which have a harder spectrum and the slow variability predominantly comes from the
outer regions which have a softer spectrum. Also, a greater number of photons emitted
from the outer regions of the flow will reflect off the disc than that from the inner
regions meaning that we should expect a greater reflection fraction for the slow
variability than for the fast variability. This effect is also observed: Revnivtsev,
Gilfanov \& Churazov (1999) show that the SED of the slow variability displays
much stronger reflection features than that of the fast variability. The frequency
dependent time lags between hard and soft energy bands (with hard lagging soft;
Miyamoto \& Kitamoto 1989; Nowak et al 1999) also follow directly from the idea of
an inhomogeneous spectrum because slow variability generated in the outer regions
modulates the soft spectrum immediately but takes time to propagate down to the inner
regions where it modulates the hard spectrum.

Taking all of this into consideration, it seems apparent that the model has the
capability to explain most the spectral variability properties of BHBs. In a
future paper (Ingram, Done \& Zycki in prep), we will include an energy dependence
in order to explicitly compare the model predictions for different properties
such as the PSD, the SED, the lag spectrum and the frequency
resolved SED simultaneously. It is important to note that no other geometry can
currently explain anything like the range of different observational properties 
that it is possible to explain with the truncated disc model.

However, although we believe the \textit{trends} in best fit parameter values to
be reliable, their \textit{absolute} values should not be taken too seriously. This
is because there are a few complexities not currently included in the model. For
example, we currently effectively assume that the disc is stable which is not
true, at least in the low/hard state (Wilkinson \& Uttley 2009). Although we only
consider energies at which the Comptonized emission dominates, the disc is feeding
the flow and therefore disc variability should propagate to the flow and modulate the
hard emission. This means that the lowest frequencies in the PSD are actually being
generated in the disc and not in the flow, meaning we over predict the truncation
radius, $r_o$. The main uncertainty associated with the model is that it is unclear
exactly how the disc and flow couple together. Although the most likely truncation
mechanism is evaporation via thermal conduction (e.g. Liu, Meyer \& Meyer-Hofmeister
1997; R{\'o}{\.z}a{\'n}ska \& Czerny 2000; Mayer \& Pringle 2007), the details of this
process are still far from well understood and, in particular, numerical simulations
of a truncated disc / hot inner flow configuration are far beyond current computing
capabilities. Whatever the specific nature of the coupling, it seems very likely
that the disc will exert a torque on the flow, especially in a region where the
flow overlaps the disc, which would slow down precession. This means that $r_o$
would need to be smaller in order for the model to reproduce both the QPO \textit{and}
the broad band noise. For this reason, we see our best fit values of $r_o$ as upper
limits rather than definitive measurements.

Still, it is extremely encouraging that this model can produce an excellent fit to PSD
data whilst also having the potential to qualitatively reproduce many other properties
seen in the data.

\section{Acknowledgements}

AI acknowledges the support of an STFC studentship. AI and CD thank
Magnus Axelsson for useful conversations on the double peaked PSD, and
Chris Fragile, Kris Beckwith and Omer Blaes for physical intuition
about the MRI.


\appendix

\section{Using \textsc{propfluc}}
\label{sec:lmod}

We intend to release the model publicly as the \textsc{xspec} local model,
\textsc{propfluc}. Here we include some tips for anyone wanting to use the model.

\subsection{Data}

We use powspec from \textsc{xronos} in order to create a power spectrum from the observed
light curve. We set norm=-2, which means white noise will be subtracted and choose the
minimum lightcurve time step, which is $dt_{obs}=0.390625\times 10^{-2}s$ for \textsc{rxte} data. 
We set the number of time steps per interval to $2^{15}=32768$, meaning that the duration of
an interval is $2^{15}dt_{obs}=128s$. This means that a periodogram will be calculated for each
interval with minimum frequency $1/128Hz$ and maximum (Nyquist) frequency $1/(2dt_{obs})=128Hz$.
The number of intervals per frame should be set to maximum so that powspec averages over as
many intervals as the length of the observation allows and we use a geometric re-binning with
a constant factor of $1.045$, resulting in $150$ new bins. The resulting binned power spectrum
can then be written to a data file in the form
\begin{equation}
f,~df,~P,~dP. \nonumber
\end{equation}
\textsc{xspec}, however is expecting to recieve data in the form
\begin{equation}
E_{min},~E_{max},~F(E_{max}-E_{min}),~dF(E_{max}-E_{min})
\nonumber
\end{equation}
where $E_{min}$ and $E_{max}$ are the lower and upper bounds of each energy bin and $F$ is
the flux. It is therefore neccessary to create a data file with inputs
\begin{equation}
f-df,~f+df,~2Pdf,~2dPdf.
\nonumber
\end{equation}
We then use flx2xsp in order to convert this into a $.pha$ file and also generate a
diagonal response function. The data can now be loaded into \textsc{xspec} and, eventhough
the axis on the plots are by default labeled as flux and energy, it is in fact reading
in a power spectrum as a function of temporal frequency (i.e the command ip euf will show
frequency multiplied by power plotted against frequency for both data and model).

\subsection{Model}

The model consists of a fortran program, $propfluc.f$, and a data file $lmodel\_pf.dat$. These
two files are all that is needed to load the model using the local model functionality.
The model has 18 parameters, summarised in table \ref{tab:paras}, plus \textsc{xspec}
always includes a 19th normalization parameter which must be set to (and fixed at) unity.
The simulated light curve is generated using a time step of $dt=dt_{obs}/4=9.76562
\times 10^{-4}s$. It is important that this time step is short because the Nyquist frequency
must be higher than the highest frequency at which significant variability is generated.
The final power spectrum is calculated using $2^{17}$ steps per interval, meaning that each
interval is $2^{17}dt=2^{17}dt_{obs}/4=2^{15}dt_{obs}=128s$. The simulated power spectrum is
then binned into the same frequency bins used for the observed power spectrum. For this
reason, it is vital that the periodograms are calculated on the same interval (i.e. 128s) for
both model and data, the use of two different intervals could result in empty bins in
the simulated power spectrum which doesn't help $\chi^2$! In table \ref{tab:paras} we see
that it is possible for the user to decide on the length of simulated light curve
(parameter 17). Since the interval length is fixed, this dictates how many intervals the power
spectrum is averaged over. We recommend $nn=22$ (32 intervals) for fitting but this does make
the code very slow. Preliminary fitting is best done with $nn=20$ (8 intervals) as this is faster
but provides a good enough PSD estimate to work with. It should be noted that this setting
slightly under predicts the power but it is a constant offset and so the best fit found
using $nn=20$ has a higher value for $F_{var}$ than that found using $nn=22$ but the
other parameters are largely unaffected. The main advantage of using $nn=22$ is that $\chi^2$
gives a much more reliable estimate of goodness of fit.

The model is difficult to fit, partly because of the stochastic nature of the power
spectrum and partly because of the complicated relationship between parameters. We recommend
finding a good fit by eye first and fixing a few key parameters before fitting. We set \textsc{xspec}
to calculate the gradient in $\chi^2$ numerically rather than analytically and set the critical
$\Delta \chi^2$ value to $0.1$ rather than the default $0.01$. Finally, the third column of table
\ref{tab:paras} shows all of our best fit model parameters for observation 1, with a $\equiv$ symbol
indicating that the parameter is fixed.

\begin{table*}
\begin{tabular}{l|l|l|l|c}
 \hline

 & Parameter & Comments & Value for obs 1

 \\
 \hline
 \hline
1  & Sigma0     & Normalization of surface density.                                        & 5.43  \\
 \hline
2  & rbw        & Bending wave radius - dictates where $\Sigma(r)$ breaks.                 & 4.60 \\
 \hline
3  & kappa      & Dictates sharpness of the break.                                         & $\equiv 3.0$ \\
 \hline
4  & lambda     & Dictates $\Sigma(r)$ for $r<r_{bw}$.                                     & $\equiv 0.9$  \\
 \hline
5  & zeta       & Dictates $\Sigma(r)$ for $r>r_{bw}$.                                     & $\equiv 0.0$  \\
 \hline
6  & Fvar       & Intrinsic amount of variability generated per decade in $f_{visc}$.      & 0.32 \\
 \hline
7  & fbmin      & This is $f_{visc}(r_o)$. It is much easier to set this instead of $r_o$. & 0.129 \\
 \hline
8  & ri         & Inner radius                                                             & $\equiv 3.3$ \\
 \hline
9  & sig\_qpo    & QPO width (fundamental). Width of higher harmonics is tied to this.      & 0.0226 \\
 \hline
10 & sig\_subh   & Width of the subharmonic. This can have a different $Q$ value to the    &   \\
        & &      other harmonics. & 0.0283  \\
 \hline
11 & n\_qpo      & Normalization of fundamental (first harmonic).                           & 0.244806 \\
 \hline
12 & n\_h        & Normalization of second harmonic.                                        & 0.1706   \\
 \hline
13 & n\_3h       & Normalization of third harmonic.                                         & 0.1018   \\
 \hline
14 & n\_subh     & Normalization of sub-harmonic.                                           & 0.0967 \\
 \hline
15 & em\_in      & Emissivity index (i.e. $\gamma$ in the text).                            & 5.281  \\
 \hline
16 & dL         & The model gives the option to generate a Gaussian error on each point of &  \\
   &            & the simulated lightcurve, thus creating white noise. To match a typical \textsc{rxte} & $\equiv 0.0$ \\
	& &	  observation, this needs to be $dL \sim 0.8$, however we recommend setting  & \\
	& &	  this to zero and using white noise subtracted data.                      &  \\
 \hline
17 & nn         & Sets the number of time steps in the simulated light curve (i.e. the light curve has &  \\
        & &          a total duration of $2^{nn}dt$). This must be an integer because the model & $\equiv 22.0$ \\
	& &	  uses a fast fourier transform algorithm (Press et al 1992). The PSD estimate of &  \\
	& &	  the model must be calculated on the same interval as the data (128s) and  &  \\
	& &	  therefore the value of nn used dictates how many intervals are averaged over. &  \\
 \hline
18 & Ndec       & Sets the radial resolution. If this is particularly high, the code is very  & \\
           & &    slow! $N_{dec}=15$ should be sufficient. The total number of annuli used is & $\equiv 15.0$ \\
	   & &    calculated from this.                             &  \\

\hline
\end{tabular}
\caption{Summary of model parameters.}

\label{tab:paras}
\end{table*}

\label{lastpage}

\end{document}